\documentclass{article}

\begin{document}

\title{Resonant pumping in nonlinear Klein-Gordon equation and solitary packets of waves
\thanks{This work was supported by grants RFBR 03-01-00716,
Leading Scientific Schools 1446.2003.1 and INTAS 03-51-4286.}}

\author{Oleg Kiselev\thanks{Institute of Math. USC
RAS; ok@ufanet.ru}, Sergei Glebov\thanks{Ufa State Petroleum Technical University; glebskie@rusoil.net} , Vladimir Lazarev\thanks{Ufa State Petroleum
Technical University; lazva@mail.ru}}

\date{}

\maketitle

\begin{abstract}
Solution of the nonlinear Klein-Gordon equation perturbed by small
external force is investigated. The frequency of perturbation varies
slowly and passes through a resonance. The
 resonance generates a solitary packets of waves.
  Full asymptotic description of this process is presented.
\end{abstract}

\font\Sets=msbm10
\def\Real{\hbox{\Sets R}}
\def\Complex{\hbox{\Sets C}}
\def\bb{\begin{equation}}
\def\ee{\end{equation}}
\def\pt{\partial}
\def\mod{\hbox{mod}}
\def\const{\hbox{const}}
\def\sgn{\hbox{sgn}}
\def\Arg{\hbox{Arg}}
\def\a{\alpha}
\def\b{\beta}
\def\d{\delta}
\def\G{\Gamma}
\def\g{\gamma}
\def\e{\epsilon}
\def\ve{\varepsilon}
\def\k{\kappa}
\def\l{\lambda}
\def\O{\Omega}
\def\o{\omega}
\def\th{t}
\def\s{\sigma}
\def\t{\tau}
\def\z{\zeta}
\newtheorem{lemma}{\bf Lemma}
\newtheorem{theorema}{\bf Theorem}

\noindent{\bf Introduction}
\par
This work is devoted to the problem  on a generation of solitary
packets of waves  by a small external driving force. We propose a
new approach for generation of solitary packets of waves. We
demonstrate that for perturbed nonlinear Klein-Gordon equation.
\par
In our approach the wave packets appear due to passing of
external driving force through resonance. After the resonance the
envelope function of the wave packet is determined by nonlinear
Schr\"odinger equation (NLSE). In the most important cases the
envelope function is a sequence of solitary waves which are
called  solitons. The wave packets with the solitons as the
envelope function  are propagated without a deformation.  The
 parameters of the solitons are obviously defined by the value of the
driving force on a resonance curve.
\par
Here we give the mathematical basis for the proposed approach.
This basis allows to derive  explicit formulas which define
parameters for the solitary packets of waves with respect to the
external driving force.  Generation of the solitary packets of
waves by a small driving force  is described in detail. The
 formulas for the asymptotic solution before, after
 and in the neighborhood of the resonance curve are obtained.
 \par
Proposed approach is based on  a local resonance phenomenon. The
local resonance in linear ordinary differential equations was
investigated in papers \cite{Kevorkyan,Rozenfeld}. Later this
phenomenon was investigated in partial differential equations in
linear case \cite{Neu} and in weak nonlinear case
\cite{Kalyakin1,Glebov}. As was shown in these papers the amplitude
of the wave which  crosses the local resonance increases by linear
way. The increase of the amplitude is proportional to the width of
the local resonance layer.
\par
After resonance  a special proportion between the order of solution
and scales of independent variables appears. This magic proportion
gives NLSE for envelope function. The deriving of NLSE in such case
is well known \cite{Kelley,Talanov,Zakharov} and  justified
\cite{Kalyakin0}.
\par
Important kind of solution of NLSE is solitons. It is known the
phenomenon of solitary waves generation for some nonlinear
equations due to the modulation instability
\cite{Kadomtsev-Petviashvili}. For example,  there exist the
detailed analytical description of disappearance (generation) of
soliton due to the modulation instability in the case of
Kadomtsev-Petviashvili equation \cite{TeoriyaSolitonov}. Some
results on soliton appearance in nonlinear Schrodinger equation
due to instability are presented in
\cite{Alexeeva-Barashenkov-Pelinovsky}. It is known the
structural instability of the solitons for Davey-Stewartson
equation \cite{grr-ok}. Perturbations used the different kinds of
instability of the solution do not allow to obtain the solitons
with  given parameters.
\par
The resonance generation of solitary waves  by small external
force is known due to computer simulation
\cite{Friedland-Shagalov}.  Although computer simulations justify
the possibility of soliton generation by an external driving
force but also do not allow to connect the the parameters of the
soliton and perturbation. Therefore the problem on soliton
generation with the given parameters was still open.
\par
The goal of this paper is the following: to demonstrate that the
process of solitary waves generation due to local resonance is
universal. This process allows to control parameters of generated
waves. Earlier this phenomenon in the case of nonlinear Schrodinger
equation was asymptotically investigated in
\cite{Glebov-Kiselev-Lazarev}.  In this work we consider the similar
phenomenon in the nonlinear Klein-Gordon equation. Our approach
demonstrates that  solitary waves with the given parameters can be
obtained for nonlinear wave systems.
\par
This paper has the following structure. The first section contains
the main result and example. The second section contains the
asymptotic construction in the pre-resonance domain. In the third
section we construct  the asymptotic solution in the neighborhood of
the resonance curve. The fourth section of the paper is devoted to
construction of the post-resonance asymptotics. All asymptotics are
matched.
\par

\section{Main result}

\par
Let us consider the Klein-Gordon equation with cubic nonlinearity
\bb
\pt^2_{t} U - \pt^2_{x} U + U + \gamma U^3 = \ve^2 f(\ve
x)\exp\left\{i {{S(\ve^2t,\ve^2x)}\over
{\ve^2}}\right\}+\hbox{c.c.},\quad 0 < \ve \ll 1. \label{we}
\ee
Here $\gamma=\const$;  $f(y)$ is smooth and rapidly vanish as
$y\to\pm\infty$. The phase function $S(y,z)$ of the driving force
and all derivates with respect to $y,z$ are bounded.
\par
We construct the formal asymptotic solution of the WKB-type using the
combination of method of multiple scales \cite{JK} and matching
method \cite{Il'in}. Below we use scaled variables
$$
x_j = \ve^j x,\quad t_j = \ve^j t, \quad j=1,2.
$$
\par
The leading-order term of the asymptotic solution has an order
$\ve^2$ and oscillates with frequency of the driving force. The
resonance curve is determined by
$$
l(t_2,x_2)\equiv (\pt_{t_2}S)^2-(\pt_{x_2}S)^2-1=0.
$$
We assume that the curve $S=const$ does not touch the resonance
curve:
$$
\pt_{x_2}l\pt_{x_2}S-\pt_{t_2}l \pt_{t_2}S\not=0.
$$
The frequency of the forced oscillations and  frequency of the eigen
oscillations of the linearized Klein-Gordon equation are equal on
the resonance curve. It yields the local resonance layer in the
neighborhood of the curve $l(x_2,t_2)=0$. The asymptotics of the
WKB-type is not valid into this layer. The leading-order term of the
asymptotics is defined by Fresnel integral. After resonance the
solution has an order $\ve$ and oscillates. The envelope function
 satisfies NLSE.
\par
The accurate formulation of the result for this paper is following
\begin{theorema}\label{mainTheorem}
In the domain $-l \gg \ve$ the asymptotic solution of (\ref{we})
modulo $O(\ve^{N+1})$ has the form
$$
U=\sum_{n\ge 2}^N \ve^n \stackrel{n}{U}(t,x,\ve),
$$
where the leading-order term is
$$
\stackrel{2}{U} = - {f \over l}\exp(iS(t_2,x_2)/\ve^2)+c.c.
$$
The higher-order terms are determined from  the recurrent system of
algebraic equations (\ref{2th}), (\ref{3th}), (\ref{4th}),
(\ref{nth}).
\newline
In the domain $|l|\ll1$ the asymptotic solution of (\ref{we})
modulo $O(\ve^{N+1})$ has the form
$$
U=\sum_{n\ge 1}^N \ve^n \stackrel{n}{W}(t_1,x_1,t_2,x_2,\ve).
$$
The leading-order term is
$$
\stackrel{1}{W}(t_1,x_1,t_2,x_2,\ve)=\stackrel{1}{W}_1\exp(iS(
t_2,x_2)/\ve^2)+c.c.
$$
The function  $\stackrel{1}{W}_1$  is solution of the equation
$$
2i\pt_{t_2}S\pt_{t_1}\stackrel{1}{W}_1 - 2i\pt_{x_2}S \pt_{x_1}
\stackrel{1}{W}_1 - \lambda \stackrel{1}{W}_1 = f
$$
with the zero condition as $\l\to-\infty$, where
$\l(x_1,t_1,\ve)=l(t_2,x_2)/\ve$. The higher-order
 terms  are either solutions of the
problems for equations (\ref{eqForN1-thOrderTerm}) with zero
conditions as $\l\to-\infty$ or solutions of the algebraic equations
(\ref{eqForNk-thCorrectionTerm}).
\newline
In the domain $l \gg \ve$ the asymptotic solution of (\ref{we})
modulo $O(\ve^{N+1})$ has the form
$$
U(x,t,\ve)=\sum_{n=1}^N\ve^n \sum_{k=0}^{n-2} \ln^k(\ve)
\bigg(\sum_{\pm\Phi}\exp\{\pm i\Phi(x_2,t_2)/\ve^2\}
\stackrel{n,k}{\Psi}{}_{\pm\Phi}(x_1,t_1,t_2)+
$$
$$
+ \sum_{\chi \in K'_{n,k}}\exp\{i\chi(x_2,t_2)/\ve^2\}
\stackrel{n,k}{\Psi}{}_{\chi}(x_1,t_1,t_2)\bigg).
$$
The phase function  $\Phi$ satisfies eikonal equation
$$
(\pt_{t_2}\Phi)^2 - (\pt_{x_2}\Phi)^2 -1 =0
$$
with conditions
$$
\Phi|_{l=0} = S|_{l=0}, \quad \pt_{t_2}\Phi|_{l=0} =
\pt_{t_2}S|_{l=0}.
$$
The envelope function of the leading-order term is a solution of the
nonlinear Schrodinger equation
$$
2i\pt_{t_2}\Phi\pt_{t_2}\stackrel{1,0}{\Psi}{}_{\Phi} +
\pt^2_{\xi}\stackrel{1,0}{\Psi}{}_{\Phi} +  i[\pt_{t_2}^2\Phi -
\pt_{x_2}^2\Phi]\stackrel{1,0}{\Psi}{}_{\Phi} + \gamma
|\stackrel{1,0}{\Psi}{}_\Phi|^2 \stackrel{1,0}{\Psi}{}_{\Phi} =0,
$$
where the $\xi$ is defined from
$$
{{dx_1}\over{d\xi}}=\pt_{t_2}\Phi,\quad
{{dt_1}\over{d\xi}}=\pt_{x_2}\Phi.
$$
The initial condition for $\stackrel{1,0}{\Psi}{}_{\Phi}$ is
$$
\stackrel{1,0}{\Psi}{}_{\Phi}|_{l=0}= \int_{-\infty}^{\infty} d\s
f(x_1)\exp(i\int_0^\s d\mu \l(x_1,t_1,\ve)),
$$
Integration in this integral is realized in the line of
characteristic direction connected with (\ref{eqForCharacteristics}),
(\ref{initialConditionsForCharacteristics}). The coefficient
$\stackrel{n,k}{\Psi}{}_{\Phi}$ is determined from Cauchy problems
for linearized Schrodinger equation (\ref{lSh}). The coefficients
$\stackrel{n,k}{\Psi}_{\chi}$, $\chi\in K'_{n,k}$ are determined
from algebraic equations  (\ref{algebraic}). The set $K_{n,k}$ is
defined by
$$
K_{1,0}={\pm\Phi};\quad K_{2,0}={\pm\Phi,\pm S},\quad
K_{3,1}={\pm\Phi},\quad K_{n,k}=\cup \alpha+\beta+\delta,
$$
where
$$
\alpha \in K_{j_1,l_1},\ \beta \in K_{j_2,l_2},\ \delta \in
K_{j_3,l_3},\ j_1+j_2+j_3=n,\ l_1+l_2+l_3=k
$$
The set $K'_{n,k}=K_{n,k}\backslash \{\pm \Phi\}$.
\end{theorema}
\par
This theorem is a direct consequence of theorems 2,3 and 4 which are
proved in the next parts of the paper.
\par
To illustrate the theorem we consider equation (\ref{we}) with the
simplest of driving force. Let the phase function of the driven be
$S=(\ve^2 t)^2/2$. In this case the curve of the local resonance is
the line $t_2=1$. In the domain $t_2
> 1$  the leading-order term of the asymptotics satisfies the Cauchy
problem
$$
2i\pt_{t_2}\stackrel{1,0}{\Psi}{}_{\Phi} +
\pt^2_{\xi\xi}\stackrel{1,0}{\Psi}{}_{\Phi} + \gamma
|\stackrel{1,0}{\Psi}{}_{\Phi}|^2 \stackrel{1,0}{\Psi}{}_{\Phi}
=0,
$$
$$
\stackrel{1,0}{\Psi}{}_{\Phi}|_{t_2=1}= f(\xi)(1+i)\sqrt{\pi}.
$$
The solution of this Cauchy problem contains solitary waves if
the initial data is sufficiently large \cite{TeoriyaSolitonov}.
\par

{\bf Remark on WKB asymptotics.} Theorem \ref{mainTheorem} describes
the special asymptotic solution of equation (\ref{we}). It is
defined by the  driving force. One can add any solution of the
WKB-type \cite{JK} of the order $\ve^2$ to this constructed
solution. It leads to an asymptotic solution for equation (\ref{we})
of the form
$$
\tilde U=U(t,x,\ve)+ \sum_{n\ge 2}^N \ve^n \stackrel{n}{\bf
U}(t,x,\ve).
$$
The coefficients $\stackrel{n}{\bf U}(t,x,\ve)$ of the asymptotics
are calculated by standard methods of WKB-theory. This additional
term leads to  ponderous formulas and does not change  the
leading-order term of the asymptotics constructed in theorem
\ref{mainTheorem}.
\par

\section{Pre-resonance expansion}
\label{externalAsymptotics1}
\par
In this section the formal asymptotic solution is constructed in the
domain before resonance. The asymptotic expansion has the form of
the WKB-type. The leading-order term of the asymptotics has the order
of the driving force and oscillates with its frequency.  The
constructed asymptotics is valid as $-l\gg\ve$. The result of this
section  is formulated below.
\par
Let us construct the formal asymptotic solution for equation
(\ref{we})  in the form
\bb
U=\sum_{n\ge 2}^N \ve^n \stackrel{n}{U}(t,x,\ve), \label{ext1}
\ee
where
$$
\stackrel{n}{U}=\sum_{k \in \Omega_n} \stackrel{n}{U}_k(t_2,x_2,\ve
x)\exp\left\{i k {{S(t_2,x_2)}\over {\ve^2}}\right\}.
$$
Set $\Omega_n$ for the higher-order  term with the number $n$ is
described by the formula
$$
\Omega_n = \Bigg\{ \begin{array}{c} \{\pm 1\}, \quad n \le 5;\\
\{\pm 1,\pm 3,\dots,\pm (2l+3)\}, \quad l=\big[(n-6)/4\big], \quad
n\ge 6.
\end{array}
$$
The functions $\stackrel{n}{U}_k$ and $\stackrel{n}{U}_{-k}$ are
complex conjugated.
\par
Let us substitute (\ref{ext1}) in equation (\ref{we}) and collect
the terms of the same order of $\ve$. As a result we obtain a
recurrent sequence of algebraic equations.
\bb
\stackrel{2}{U}_{1} = - {f \over l}, \label{2th}
\ee
\bb
\stackrel{3}{U}_{1} = 2i{{\pt_{x_1}f \pt_{x_2}S} \over l^2},
\label{3th}
\ee
\begin{eqnarray}
\stackrel{4}{U}_{1} = {{2if[\pt_{t_2}S \pt_{t_2}l -
\pt_{x_2}S\pt_{x_2}l] -4( \pt_{x_2}S)^2 \pt^2_{x_1}f } \over l^3} -
\nonumber\\
{{2i \pt_{t_2}f\pt_{t_2}S + \pt^2_{x_1}f + i \pt^2_{t_2}S f} \over
l^2}, \label{4th}
\end{eqnarray}
where
$$
l=(\pt_{t_2}S)^2 - (\pt_{x_2}S)^2-1.
$$
\par
The curve where the phase function $S$ satisfies eikonal equation is
called the  resonance curve
\bb
l[S]=(\pt_{t_2}S)^2 - (\pt_{x_2}S)^2-1=0. \label{eikonal}
\ee
The amplitude $\stackrel{n}{U}_{1}$ has a singularity on this curve.
\par
The formula for the $n$-th order term has the form
$$
\stackrel{n}{U}_k=\frac1l\Big[\partial^2_{t_2t_2}\stackrel{n-4}{U}_{\!\!k}+
2ik\pt_{t_2}S\partial_{t_2}\stackrel{n-2}{U}_{\!\! k} +
ikS_{t_2t_2}\stackrel{n-2}{U}_{\!\!k}-
2ik\pt_{x_2}S\partial_{x_2}\stackrel{n-2}{U}_{\!\! k}-
ik\pt^2_{x_2}S\stackrel{n-2}{U}_{\!\!k} -
$$
\bb
\partial^2_{x_1x_1}\stackrel{n-2}{U}_{\!\!k}-
2\partial^2_{x_1x_2}\stackrel{n-3}{U}_{\!\!k}-
\partial^2_{x_2x_2}\stackrel{n-4}{U}_{\!\!k}-
2ik\pt_{x_2}S\partial_{x_1}\stackrel{n-1}{U}_{\!\!k}+\gamma
\sum_{ \begin{array}{c}{\scriptstyle n_1+n_2+n_3=n }, \\
{\scriptstyle k_1 + k_2 +k_3 =k }\\{\scriptstyle k\in \Omega_{n}}
\end{array}} \hskip-0.3cm \stackrel{n_1}{U}_{k_1}
\stackrel{n_2}{U}_{k_2} \stackrel{n_3}{U}_{k_3} \Big]. \label{nth}
\ee
\par
\begin{lemma} The coefficient $\stackrel{n}{U}_{k}$ has the following
behaviour
\bb
\stackrel{n}{U}_{k} = O(l^{-(n-k)}),\ k >0 ,\qquad l\to -0.
\label{syngord}
\ee
\end{lemma}
{\bf Proof.} Let us prove this lemma as $k=1$.  The validity of
formula (\ref{syngord}) for  $n=2,3,4$ directly obtains from
(\ref{2th}), (\ref{3th}), (\ref{4th}). Suppose now that this formula
is valid for the term $\stackrel{n-1}{U}_{1}$. The increase of the
order of the singularity as $l\to 0$ takes place due to
differentiation with respect to $x_2, t_2$ and the nonlinear term in
formula (\ref{nth}). Differentiation of the terms in formula
(\ref{nth}) leads to formula (\ref{syngord}).
\par
Let us consider $\stackrel{n}{U}_{k}$ for $k>1$.
 The validity of formula (\ref{syngord}) for small
values of $n$ and $k$ obtains by direct calculations.  Consider the
$n-$th order term. It contain the terms with different values of
$k$. The higher-order term with $k=3$ have the greatest order of
singularity.
\bb
\stackrel{n}{U}_{3} = O(l^{-(n-3)}),\quad l\to -0. \label{ksyng}
\ee
It takes place because the right hand side of (\ref{nth}) contains
the term $\stackrel{n-4}{U}\!\!{}_{\pm 1}\stackrel{2}{U}_{\pm
1}\stackrel{2}{U}_{\pm 1}$. The calculation of the order of
singularity  for this term leads to formula (\ref{ksyng}). The terms
of the type of $\stackrel{n_3}{U}\!\!{}_{\pm
3}\stackrel{n_1}{U}_{\mp k_1}\stackrel{n_2}{U}_{\pm k_1},\
n_1+n_2+n_3 =n$ lead to weak singularities, for example for $k_1=3$
we obtain the order of singularity is equal to $n-9$.
\par
 Consider
nonlinear term $\stackrel{n_1}{U}_{k_1}\stackrel{n_2}{U}_{k_2}
\stackrel{n_3}{U}_{k_3}$ from right hand side of (\ref{nth}) when
the number of the higher-order term is equal to $n$. Calculate the
order of singularity for this term using the $(n-1)-$th step of
induction. Indexes of the amplitudes are connected by formulas
$$
n_1 + n_2 + n_3 = n,\quad k_1 + k_2 + k_3 = k.
$$
Using (\ref{syngord}) for $n_1, n_2, n_3 < n$ we obtain that the
order of the singularity for this term is equal to $(n-k)$.
\par
The right hand side of (\ref{nth}) contains derivatives of previous
terms with respect to $x_2, t_2$. It leads to increase of the order
of the singularity but the leading order nevertheless we obtain from
nonlinear terms. The lemma is proved.
\par
The domain of validity as $l \to -0$ for formal asymptotic solution
in the form (\ref{ext1}) is defined by
$$
{{\ve^{n+1}\stackrel{n+1}{U}}\over {\ve^{n}\stackrel{n}{U}}} \ll 1.
$$
It yields
$$
-l \gg \ve.
$$
Using these lemmas we obtain the asymptotic representation for
(\ref{ext1}) as $l \to -0$
\bb
U = \sum_{n=2}^N \ve^n \sum_{k \in \Omega_n} \exp\{ikS/\ve^2\} \sum_{j=-(n-k)}^{\infty}\stackrel{n}{U}\!\!{}^j_{k}\ l^j,\quad l \to -0. \label{externalAsymptoticsCloseToSingularity}
\ee
The following theorem is proved.
\par
\begin{theorema}
In the domain $-l \gg \ve$ the formal asymptotic solution of
equation (\ref{we}) modulo $O(\ve^{N+1})$ has the form
(\ref{ext1}). The coefficients of the asymptotics
$\stackrel{n+1}{U}\!\!{}_k$ are defined from algebraic equations
(\ref{2th}), (\ref{3th}), (\ref{4th}), (\ref{nth}).
\end{theorema}

\section{Internal asymptotics}\label{internalAsymptotics}

This part of the paper contains the asymptotic construction of
the solution for equation (\ref{we}) in the neighborhood of the
curve $l=0$. The domain of validity of this asymptotics
intersects with domain of validity of expansion (\ref{ext1}).
These expansions are matched.
\par
\begin{theorema}\label{internalAsymptoticTheorem}
In the domain $|l|\ll1$ the formal asymptotic solution for
equation (\ref{we}) modulo $O(\ve^{N+1})$ has the form
\bb
U=\sum_{n\ge 1}^N \ve^n \stackrel{n}{W}(t_1,x_1,t_2,x_2,\ve),
\label{int1}
\ee
where
\bb
\stackrel{n}{W}=\sum_{k \in \Omega_n}
\stackrel{n}{W}_k(x_2,t_2,x_1,t_1)\exp\left\{i k {{S(t_2,x_2)}\over
{\ve^2}}\right\}, \label{int2}
\ee
The function $\stackrel{n}{W}_k,\ k=1$ is solution of the problem
for equation (\ref{eqForN1-thOrderTerm}) with zero condition as
$\l\to-\infty$ and solutions of algebraic equations
(\ref{eqForNk-thCorrectionTerm}) in the case $k\not= 1$. The
functions $\stackrel{n}{W}_k$ and $\stackrel{n}{W}_{-k}$ are complex
conjugated.
\end{theorema}
\par
There is an essential difference between asymptotics (\ref{int1}) and
external pre-resonance asymptotics (\ref{ext1}). First the leading
order term in (\ref{int1}) has an order $\ve$ in contrast the
leading order term in (\ref{ext1}) has an order $\ve^2$. Second the
coefficients of asymptotics (\ref{int1}) depend on fast variables
$x_1=x_2/\ve$ and $t_1=t_2/\ve$.
\par
The proof of theorem \ref{internalAsymptoticTheorem} consists in
three steps. First we derive  equations for coefficients of the
asymptotics. Second we solve the problems for asymptotic
coefficients. And third we determine the domain of the validity for
expansion (\ref{int1}).
\par

\subsection{The equations for coefficients}
\par
Let us construct the internal asymptotic expansion in the domain
 $|l|\ll 1$. Denote
\bb
\l(x_1,t_1,\ve)={1\over \ve}l(\ve x_1,\ve
t_1).\label{definitionOfTheLambda}
\ee
\par
In the domain $1 \ll \l \ll \ve^{-1}$ both asymptotics (\ref{ext1})
and (\ref{int1}) are valid. This fact allows us to obtain the
asymptotic representation for coefficients of the internal
asymptotics. Substitute $l=\ve\l$ in formula
(\ref{externalAsymptoticsCloseToSingularity}) and expand the
obtained expression with respect to powers of small parameter $\ve$.
It yields
\bb
\stackrel{n}{W}_{k}=\sum_{j=(n-k+1)}^\infty\l^{-j}\stackrel{n+1}{U}{}^j_{k}(x_2,t_2,x_1),
\quad k\in \Omega_n,\quad \l \to -\infty.
\label{internalAsymptoticsCloseTo-Infinity}
\ee
\par
Let us obtain the differential equations for the coefficients of
asymptotics (\ref{int1}). Substitute (\ref{int1}),
 (\ref{int2}) in equation (\ref{we}) and collect the terms with
equal powers of small parameter and exponents. It yields the
equations for coefficients $\stackrel{n}{W}_k$. In particularly, the
terms of the order $\ve^2$ give us the equations for the
leading-order terms of the asympotics
\bb
2i\pt_{t_2}S\pt_{t_1}\stackrel{1}{W}_1 - 2i\pt_{x_2}S \pt_{x_1}
\stackrel{1}{W}_1 - \lambda \stackrel{1}{W}_1 = f,
\label{eqForLeadingOrderTerm}
\ee
and complex conjugated equation for $\stackrel{1}{W}_{-1}$.
\par
The relation of the order $\ve^3$ in equation (\ref{we}) gives four
equations. Two of them are complex conjugate differential equations
for $\stackrel{2}{W}_1$ and $\stackrel{2}{W}_{-1}$:
\begin{eqnarray}
2i\pt_{t_2}S\pt_{t_1}\stackrel{2}{W}_1 - 2i\pt_{x_2}S
\pt_{x_1}\stackrel{2}{W}_1- \lambda \stackrel{2}{W}_1 =
\pt^2_{x_1}\stackrel{1}{W}_1-\pt^2_{t_1}\stackrel{1}{W}_1  -
\nonumber
\\ -i [\pt_{t_2}^2S - \pt_{x_2}^2S] \stackrel{1}{W}_1  -2i\pt_{t_2}S\pt_{t_2}\stackrel{1}{W}_1
+2i\pt_{x_2}S\pt_{x_2}\stackrel{1}{W}_1
 - 3\gamma |\stackrel{1}{W}_1|^2 \stackrel{1}{W}_1,\label{eqForFirstCorrectionTerm}
\end{eqnarray}
two another equations are algebraic. These last equations allow us to
determine the functions
 $\stackrel{3}{W}_3$ and $\stackrel{3}{W}{}_{-3}$
$$
\stackrel{3}{W}_3 = \frac{\gamma}{8}(\stackrel{1}{W}_1)^3.
$$
\par
The higher-order terms  are calculated by the same way. In
particularly, the  terms in the case lower index is equal to $1$ are
determined by differential equations.
\bb
2i\pt_{t_2}S\pt_{t_1}\stackrel{n}{W}_1 - 2i\pt_{x_2}S \pt_{x_1}
\stackrel{n}{W}_1 - \lambda \stackrel{n}{W}_1 = \stackrel{n}{F}_1.
\label{eqForN1-thOrderTerm}
\ee
The right hand side of equation (\ref{eqForN1-thOrderTerm}) has the
form
\begin{eqnarray}
\stackrel{n}{F}_1= -2i\pt_{t_2}S\pt_{t_2}\stackrel{n-1}{W}_1 +
2i\pt_{x_2}S\pt_{x_2}\stackrel{n-1}{W}_1+
(\pt_{t_2}S)^2\stackrel{n-1}{W}_1 -
(\pt_{x_2}S)^2\stackrel{n-1}{W}_1-\nonumber\\
-\pt_{t_1}^2\stackrel{n-1}{W}_1 +\pt_{x_1}^2\stackrel{n-1}{W}_1-
\pt_{t_2}\pt_{t_1}\stackrel{n-2}{W}_1+
\pt_{x_2}\pt_{x_1}\stackrel{n-2}{W}_1-\nonumber \\
- \pt_{t_2}^2\stackrel{n-3}{W}_1 + \pt_{x_2}^2\stackrel{n-3}{W}_1-
\g\sum_{\begin{array}{c} n_1+n_2+n_3=n+1 , \\
 k_1 + k_2 +k_3 =1\\ k_j\in \Omega_{n_j},\,j=1,2,3 \end{array}}\stackrel{n_1}{W}_{k_1}
\stackrel{n_2}{W}_{k_2}
\stackrel{n_3}{W}_{k_3}.\label{rightSideOfEqForN1-thCorrectionTerm}
\end{eqnarray}
The higher-order terms in the case the lower index is not equal to
$1$ are determined by algebraic equations
$$
\stackrel{n}{W}_k= \frac{\gamma}{k^2 -1}\left(
-2i\pt_{t_2}S\pt_{t_2}\stackrel{n-2}{W}_k +
2i\pt_{x_2}S\pt_{x_2}\stackrel{n-2}{W}_k+
(\pt_{t_2}S)^2\stackrel{n-2}{W}_k -
(\pt_{x_2}S)^2\stackrel{n-2}{W}_k-\right.
$$
$$
-\pt_{t_1}^2\stackrel{n-2}{W}_k +\pt_{x_1}^2\stackrel{n-2}{W}_k-
\pt_{t_2}\pt_{t_1}\stackrel{n-3}{W}_k+
\pt_{x_2}\pt_{x_1}\stackrel{n-3}{W}_k-
$$
\begin{eqnarray}
- \pt_{t_2}^2\stackrel{n-4}{W}_k + \pt_{x_2}^2\stackrel{n-4}{W}_k-
\sum_{\begin{array}{c} n_1+n_2+n_3=n+1 , \\
k_1 + k_2 +k_3 =k\\ k_j\in \Omega_{n_j},\,j=1,2,3
\end{array}}
 \stackrel{n_1}{W}_{k_1}
\stackrel{n_2}{W}_{k_2} \stackrel{n_3}{W}_{k_3}\Bigg).
\label{eqForNk-thCorrectionTerm}
\end{eqnarray}

\subsection{The solvability of equations for higher-order terms}
\par
In this  section we present the explicit form for higher-order term
$\stackrel{n}{W}_1$ and investigate the asymptotic behaviour as
 $\l\to\pm\infty$.

\subsubsection{Characteristic variables}

The function $\stackrel{n}{W}_1$ satisfies equation
(\ref{eqForN1-thOrderTerm}). The solution is constructed by
characteristic method. Define the characteristic variables $\s,\xi$.
We choose a point $(x^0_1,t^0_1)$ such that
$\pt_{x_2}l|_{(x^0_1,t^0_1)}\not=0$ as origin and denote by $\s$ the
variable along the characteristic family for equation
(\ref{eqForN1-thOrderTerm}). We suppose  $\s=0$ on the curve $\l=0$.
The variable $\xi$ mensurates the distance along the curve $\l=0$
from the point $(x^0_1,t^0_1)$. This point $(x^0_1,t^0_1)$
corresponds to $\xi=0$. Let  the positive direction for parameter
$\xi$ coincide with positive direction of $x_2$  in the neighborhood
of $(x^0_1,t^0_1)$.
\par
The characteristic equations for (\ref{eqForN1-thOrderTerm}) have a
form
\bb
{dt_1\over d\s}=2\pt_{t_2}S(\ve x_1,\ve t_1),\quad {dx_1\over
d\s}=-2\pt_{x_2}S(\ve x_1,\ve t_1).\label{eqForCharacteristics}
\ee
The initial conditions for the equations are
\bb
x_1|_{\s=0}=x^0_1,\quad
t_1|_{\s=0}=t^0_1.\label{initialConditionsForCharacteristics}
\ee
\begin{lemma}\label{lemmaAboutSolvabilityForCharateristicEq}
The Cauchy problem for characteristics has a solutions as
 $|\s|<c_1\ve^{-1},\quad c_1=const>0$.
\end{lemma}
{\bf Proof.} The Cauchy problem (\ref{eqForCharacteristics}),
(\ref{initialConditionsForCharacteristics}) is equivalent  to the
system of the integral equations
\bb
t_1=t^0_1+2\int_0^\s  \pt_{t_2}S(\ve x_1,\ve t_1)d\z,\quad
x_1=x^0_1-2\int_0^\s  \pt_{x_2}S(\ve x_1,\ve t_1)d\z.
\label{integralEqForCharacteristics}
\ee
Substitute $\tilde t_2=(t_1-t^0_1)\ve,\,\,\tilde
x_2=(x_1-x^0_1)\ve$. It yields
$$
\tilde t_2=2\int_0^{\ve \s}  \pt_{t_2}S(\tilde x_2-\ve x^0_1,\tilde
t_2-\ve t^0_1)d\z,\quad \tilde x_2=-2\int_0^{\ve\s}
 \pt_{x_2}S(\tilde x_2-\ve x^0_1,\tilde t_2-\ve t^0_1)d\z.
$$
The integrands are smooth and bounded functions on the plane
$x_2,t_2$. There exists the constant $c_1=\const>0$ such that the
integral operator is contraction operator as $\ve|\s|<c_1$. Lemma
 \ref{lemmaAboutSolvabilityForCharateristicEq} is proved.
\par
{\bf Assumption.} We assume that the change of variables
$(x_1,t_1)\to(\s,\xi)$ is unique in the neighborhood of the curve
$\l=0$. This assumption means that the characteristics for equation
(\ref{eqForN1-thOrderTerm}) do not touch the curve $\l=0$. It means
$$
\pt_{x_2}l\pt_{x_2}S-\pt_{t_2}l \pt_{t_2}S\not=0.
$$
\par
It is convenient to use the following asymptotic formulas for change
of variables  $(x_1,t_1)\to(\s,\xi)$.

\begin{lemma}\label{lemmaAboutAsymptoticsForCharacteristics}
In the domain $|\s|\ll\ve^{-1}$ the asymptotics  as $\ve\to0$ of the
solutions for Cauchy problem (\ref{eqForCharacteristics}),
(\ref{initialConditionsForCharacteristics})  have the form
\begin{eqnarray}
x_1(\s,\xi,\ve)-x^0_1(\xi)=-2\s \pt_{x_2}S+2\sum_{n=1}^N
\ve^n\s^{n+1} g_n(\ve x_1,\ve t_1)+O(\ve^{N+1}\s^{N+2}),\qquad
\label{asymptoticsOf-x1}\\
t_1(\s,\xi,\ve)-t^0_1(\xi)=2\s \pt_{t_2}S+2\sum_{n=1}^N
\ve^n\s^{n+1} h_n(\ve x_1,\ve t_1)+O(\ve^{N+1}\s^{N+2}),\qquad
\label{asymptoticsOf-t1}
\end{eqnarray}
where
$$
g_n=-{d^n \over d \s^n}(\pt_{x_2}S)\bigg|_{\s=0},\quad  h_n={d^n
\over d \s^n}(\pt_{t_2}S)\bigg|_{\s=0}.
$$
\end{lemma}
\par
The lemma proves by integration by parts of equations
(\ref{integralEqForCharacteristics}).
\par
The next proposition gives us the asymptotic formula which connects
variables $\s$ and $\l$ as $\s,\l\to \pm\infty$.

\begin{lemma}\label{lemma_lambda_and_sigma}

Let be $\s\ll \ve^{-1}$, then:
$$
\l=\phi(\xi)\s+O(\ve\s^2),\quad \quad \phi(\xi)={d\l\over
d\s}\bigg|_{\s=0}\quad \s\to \infty.
$$
\end{lemma}
\par
{\bf Proof.} From formula (\ref{definitionOfTheLambda}) we obtain
the representation in  the form
$$
\l=\sum_{j=1}^\infty \l_j(x_1,t_1,\ve)\s^j\ve^{j-1},
$$
where
$$
\l_j(x_1,t_1,\ve)={1\over j!}{d^j\over
d\s^j}\l(x_1,t_1,\ve)|_{\s=0}.
$$
\par
It yields
$$
\l={d\l\over d\s}\big|_{\s=0} \s +O\big(\ve\s^2 {d^2 \l\over
d\s^2}\big).
$$
Let be
$$
\left|{d^2 l\over d\s^2}\right|\ge\const, \, \, \xi\in R.
$$
\par
The function $d\l/ d\s$ is not equal to  zero
$$
{d\l\over
d\s}={1\over2}\bigg(-\pt_{x_2}\l\pt_{x_2}S+\pt_{t_2}\l\pt_{t_2}S\bigg)\not=0.
$$
Let us suppose $d\l/ d\s>0$. It yields
$$
\l=\phi(\xi)\s+O(\ve\s^2),\quad \quad \phi(\xi)={d\l\over
d\s}\bigg|_{\s=0}
$$
The lemma is proved.

\subsubsection{Solutions of the equations for higher-order terms}
\par
The higher-order  terms $\stackrel{n}{W}_{\pm1}$ are solutions of
equation (\ref{eqForN1-thOrderTerm}) with the given asymptotic
behaviour $\l\to-\infty$. Equation (\ref{eqForN1-thOrderTerm}) can
be written in characteristic variables as
\bb
i{d\over d\s}\stackrel{n}{W}_1-\l
\stackrel{n}{W}_1=\stackrel{n}{F}_1.
\label{characteristicEqForN1-thOrderTerm}
\ee
\begin{lemma}\label{lemmaAboutN1-thOrderTerm}
The solution of equation (\ref{eqForN1-thOrderTerm}) with the
asymptotic behaviour (\ref{internalAsymptoticsCloseTo-Infinity})
as $\l\to-\infty$ has a form
\bb
\stackrel{n}{W}_1=\exp(-i\int_0^\s d\zeta\l(x_1,t_1,\ve))
\int_{-\infty}^{\s} d\zeta\stackrel{n}{F}_1(x_1,t_1,\ve)
\exp(-i\int_0^\z d\chi\l(x_1,t_1,\ve)). \label{N1-thOrderTerm}
\ee
\end{lemma}
\par
{\bf Proof.} By direct substitution we see that expression
(\ref{N1-thOrderTerm}) is the solution of
(\ref{characteristicEqForN1-thOrderTerm}). The asymptotics of this
solution as $\l\to-\infty$ can be obtained by integration by parts
and substitution
$$
{d\over d\s}=2\pt_{t_2}S\pt_{t_1}-2\pt_{x_2}S\pt_{x_1}.
$$
It yields
\bb
\stackrel{n}{W}_1=\sum_{j=0}^\infty
\bigg({2\pt_{t_2}S\pt_{t_1}-2\pt_{x_2}S\pt_{x_1}\over
i\l}\bigg)^j\bigg[{\stackrel{n}{F}_1\over i\l}\bigg],\quad
\l\to-\infty. \label{asymptoticsForN1-thOrderTerm}
\ee
From formula (\ref{rightSideOfEqForN1-thCorrectionTerm}) we obtain
that formulas (\ref{asymptoticsForN1-thOrderTerm}) and
(\ref{internalAsymptoticsCloseTo-Infinity}) are equivalent. The lemma
is proved.

\subsection{Asymptotics as $\l\to\infty$ and domain of validity of
the internal asymptotics}
\par
The domain of validity of the internal expansion is determined by the
asymptotics of higher-order terms. In this section we show that the
$n-$th order term of the asymptotic solution increases as
$\l^{n-1}$ when $\l\to\infty$. This increase of higher-order terms
 allows us to determine the domain of validity for
internal asymptotics (\ref{int1}) as $\l\to\infty$.
\par

\subsubsection{Asymptotics of higher-order terms}
\par
This section contains two propositions concerning asymptotic
behaviour as $\l\to\infty$ for  higher-order terms in (\ref{int1}).
The first lemma describes the asymptotic behaviour of higher-order
terms as $\l\to\infty$ and the second one contains a result about
asymptotics of the phase function.
\par
\begin{lemma}\label{lemmaAboutAsymptoticsAsLambdaToInfonity}
The asymptotic behaviour of $\stackrel{n}{W}_1$ as
$1\ll\l\ll\ve^{-1}$ has a form
\begin{eqnarray}
\stackrel{n}{W}_1=\sum_{j=0}^{n-1}\sum_{k=0}^{j-1}\bigg(\l^j\ln^{k}|\l|\stackrel{n}{W}{}\!^{(j,k)}_1(\xi)\bigg)
\exp(-i\int_0^\s d\z \l(x_1,t_1,\ve))\,+\nonumber\\
+
\sum_{j=0}^{\infty}\bigg({2\pt_{t_2}S\pt_{t_1}-2\pt_{x_2}S\pt_{x_1}\over
i\l} \bigg)^j\bigg[{\stackrel{n}{F}_1\over i\l}\bigg].
\label{asymptoticsForN1-thOrderTermAsPlusInfinity}
\end{eqnarray}
\end{lemma}
\par
{\bf Proof.} Let us calculate the asymptotics of the leading-order
term
$$
\stackrel{1}{W}_1=\exp(-i\int_0^\s d\z
\l(x_1,t_1,\ve))\int_{-\infty}^\z d\z f(x_1)\exp(i\int_0^\s d\chi
\l(x_1,t_1,\ve))=
$$
$$
\exp(-i\int_0^\s d\z \l(x_1,t_1,\ve))\int_{-\infty}^\infty d\z
f(x_1)\exp(i\int_0^\z d\chi \l(x_1,t_1,\ve)) -
$$
$$
\exp(-i\int_0^\s d\z \l(x_1,t_1,\ve))\int_{-\s}^\infty d\z
f(x_1)\exp(i\int_0^\z d\chi \l(x_1,t_1,\ve)).
$$
Further by integration by parts of the last term we obtain formula
(\ref{asymptoticsForN1-thOrderTermAsPlusInfinity}) as $n=1$, where
$$
\stackrel{1}{W}{}\!^{(0,0)}_1(\xi)=\int_{-\infty}^{\infty} d\s
f(x_1)\exp(i\int_0^\s d\chi \l(x_1,t_1,\ve)),
$$
$$
\stackrel{1}{F}_1=f(x_1).
$$
\par
To calculate the asymptotics of $\stackrel{2}{W}_1$ in formula
(\ref{N1-thOrderTerm}) we use the asymptotics with respect to
$\s$ of the leading-order term. Integral (\ref{N1-thOrderTerm})
contains the term with linear increase with respect to $\s$ when
$n=2$. We eliminate this growing part from integral explicitly.
The residuary integral converges as $\s\to\infty$. It can be
calculated in the same manner as it was calculated for
$\stackrel{1}{W}_1$. It yields formula
(\ref{asymptoticsForN1-thOrderTermAsPlusInfinity}) as $n=2$, where
$$
\stackrel{2}{W}{}\!^{(1,0)}_1(\xi)=\stackrel{1}{W}{}\!^{(0,0)}_1(\xi).
$$
The same direct calculations are realized for the $n-$th order
 term. The lemma is proved.
\par
To complete the proof of theorem \ref{internalAsymptoticTheorem} we
need to obtain the domain of validity of asymptotics (\ref{int1}).
The formal series (\ref{int1}) is asymptotic when
$$
{\ve^{n+1}\stackrel{n+1}{W}\over\ve^n\stackrel{n}{W}}\ll1,
\quad\ve\to0.
$$
Lemma \ref{lemmaAboutAsymptoticsAsLambdaToInfonity} gives
$\l\ll\ve^{-1}.$ After substitution $\l=\ve l$ we obtain $l\ll1.$
Theorem \ref{internalAsymptoticTheorem} is proved.

\subsubsection{Asymptotics of the phase function as $\l\to\infty$}
\label{internalAsymptoticsAsLambdaToInfinity}

To obtain the asymptotics as $\l\to\infty$ we need to derive the
asymptotics of the phase function in formula
(\ref{asymptoticsForN1-thOrderTermAsPlusInfinity}).
\par
\begin{lemma}\label{lemmaAboutAsymptoticsOfPhaseAsLambdaToInfinity}
As $\l\to\infty$:
\bb\label{asymptoticsOfPhaseAsLambdaToInfinity}
\int_0^\s d\xi \l={S\over\ve^2} \,+\,
{1\over\ve}(\pt_{x_2}S(x_1-x_1^0)+\pt_{t_2}S(t_1-t_1^0))+\,
O(\ve\l^3).
\ee
\end{lemma}
\par
{\bf Proof.} Substitute the asymptotics of $\l$ from lemma
\ref{lemmaAboutAsymptoticsAsLambdaToInfonity}. Calculate the
asymptotics of the integral in formula
(\ref{asymptoticsOfPhaseAsLambdaToInfinity})
$$
\int_0^\s d\z\l(x_1,t_1,\ve)=\int_0^\s
{d\z\over2}\bigg[(-\pt_{x_2}l\pt_{x_2}S+\pt_{t_2}l\pt_{t_2}S)\z\,+\,
O(\ve\z^2)\bigg]=
$$
$$
(-\pt_{x_2}l\pt_{x_2}S+\pt_{t_2}l\pt_{t_2}S){\s^2\over4}
+O(\ve\s^3).
$$
The asymptotics of the phase function $S(x_2,t_2)$ in the
neighborhood of the curve $l=0$ is represented by a segment of
the Taylor series. It yields
$$
{S\over\ve^2}={1\over\ve}(\pt_{x_2}S(x_1-x_1^0)+\pt_{t_2}S(t_1-t_1^0))+
$$
$$
{1\over2}
(S_{x_2x_2}(x_1-x_1^0)^2+2S_{x_2t_2}(x_1-x_1^0)(t_1-t_1^0)+
S_{t_2t_2}(t_1-t_1^0)^2) +
$$
$$
O(\ve(|t_1-t_1^0|+|t_1-t_1^0|)^3).
$$
Substitute instead of $(x_1-x_1^0)$ and  $(t_1-t_1^0)$ their
asymptotic behaviour with respect to $\ve$ from lemma
\ref{lemmaAboutAsymptoticsForCharacteristics}. This substitution and
result of lemma \ref{lemma_lambda_and_sigma} complete the proof of
lemma \ref{lemmaAboutAsymptoticsOfPhaseAsLambdaToInfinity}.
\par
The asymptotics as $\l\to-\infty$ contains fast oscillating terms
with phase functions $kS, k\in Z$. The leading-order term of the
asymptotics as $\l\to\infty$ contains the oscillations with an
additional phase function. We obtain this result from lemma
\ref{lemmaAboutAsymptoticsAsLambdaToInfonity}. Denote this new phase
function by $\Phi(x_2,t_2)/\ve^2$. The asymptotics of this function
is obtained in lemma
\ref{lemmaAboutAsymptoticsOfPhaseAsLambdaToInfinity}. The
nonlinearity and additional phase function lead to more complicated
structure of the phase set for higher-order terms of the asymptotics
as $\l\to\infty$.
\par
\begin{lemma}\label{lemmaAboutPhasesInAsymptoticsAsLambdaToInfinity}
The phase set $K_n$ for the $n-$th order term of the asymptotics as
$\l\to\infty$ is determined by formula
$$
K_1={\pm\Phi};\quad K_2={\pm\Phi,\pm S},\quad
K_n=\cup_{j_1+j_2+j_3=n}
\chi_{j_1}+\chi_{j_2}+\chi_{j_3},\,\,\,,\chi_{j_k} \in K_{j_k}.
$$
\end{lemma}
The proof of this lemma follows from the asymptotic formula for
$n-$th order term. Representation (\ref{int1}), formula
(\ref{asymptoticsForN1-thOrderTermAsPlusInfinity}) and lemma
\ref{lemmaAboutAsymptoticsAsLambdaToInfonity} allow us to construct
the asymptotics as $\l\to\infty$ of the internal expansion in an
explicit form
\begin{eqnarray}
U=\sum_{n=1}^N\ve^n\bigg(\sum_{j=0}^{n-1}\sum_{k=0}^{n-2}\l^j\ln^{k}|\l|\stackrel{n}{W}{}\!^{(j,k)}_1(\xi)\bigg)
\times\nonumber\\
\exp\bigg[-i\bigg({1\over\ve}(\pt_{x_2}S(x_1-x_1^0)+\pt_{t_2}S(t_1-t_1^0))+\,
O(\ve\l^3)\bigg)\bigg]\, +\,
\nonumber\\
\sum_{n=1}^N\ve^n\bigg(
\sum_{j=0}^{\infty}\bigg({2\pt_{t_2}S\pt_{t_1}-2\pt_{x_2}S\pt_{x_1}\over
i\l} \bigg)^j\bigg[{\stackrel{n}{F}_1\over
i\l}\bigg]\bigg)\exp\left\{i {{S(t_2,x_2)}\over {\ve^2}}\right\}+
\nonumber\\
\sum_{n=2}^N\ve^n\bigg(\sum_{k\in\Omega,k\not=\pm1}\stackrel{n}{W}_k\exp\left\{ik
{{S(t_2,x_2)}\over {\ve^2}}\right\}\bigg)+\,\,c.c..
\label{asymptoticOfTheSolutionAsLambdaToInfinity}
\end{eqnarray}
This representation and formula (\ref{eqForNk-thCorrectionTerm})
complete the proof of the lemma.

\section{Post-resonance expansion}
\par
This section contains the construction of the asymptotics of the
solution for (\ref{we}) after passage through resonance. The
constructed solution has the order $\ve$ and oscillates. The
envelope function of these oscillations satisfies nonlinear
Schrodinger equation. This section consists in two parts. The first
part contain the construction of the formal asymptotic solution. We
obtain the  equations for higher-order  terms of the asymptotics.
Asymptotic behaviour for higher-order terms as $l \to -0$ follows
from section \ref{internalAsymptoticsAsLambdaToInfinity}. In the
second part of this section we  determine  the domain of validity
for this external asymptotics near resonance curve $l(x_2,t_2)=0$.
The matching method gives us the initial conditions for higher-order
 terms  of the asymptotics.
\par
The main result of this section is formulated in the following
theorem.
\par
\begin{theorema} \label{theoremAboutSecondAsymptotics}
In the domain $l \gg \ve$ the formal asymptotic solution of
equation (\ref{we}) modulo $O(\ve^{N+1})$ has a form
\begin{eqnarray}
U(x,t,\ve)=\sum_1^N\ve^n \sum_{k=0}^{n-2} \ln^k(\ve)
\bigg(\sum_{\pm\Phi}\exp\{\pm
i\Phi(x_2,t_2)/\ve^2\} \stackrel{n,k}{\Psi}{}_{\pm\Phi}(x_1,t_1,t_2)+\nonumber\\
 \sum_{\chi \in
K'_{n,k}}\exp\{i\chi(x_2,t_2)/\ve^2\}
\stackrel{n,k}{\Psi}{}_{\chi}(x_1,t_1,t_2)\bigg). \label{exaf}
\end{eqnarray}
Here the function $\Phi(x_2,t_2)$ satisfies eikonal equation
\bb
(\pt_{t_2}\Phi)^2 - (\pt_{x_2}\Phi)^2 -1 =0 \label{eikonal2}
\ee
and initial condition on the curve $l=0$:
$$
\Phi|_{l=0} = S|_{l=0}, \quad
\pt_{t_2}\Phi_{l=0}=\pt_{t_2}S|_{l=0}.
$$
The leading-order term of the asymptotics is a solution of the
Cauchy problem for nonlinear Schrodinger equation
\begin{eqnarray*}
2i\pt_{t_2}\Phi\pt_{t_2}\stackrel{1,0}{\Psi}{}_{\Phi} +
\pt^2_{\xi}\stackrel{1,0}{\Psi}{}_{\Phi} +  i[\pt_{t_2}^2\Phi -
\pt_{x_2}^2\Phi]\stackrel{1,0}{\Psi}{}_{\Phi} + \gamma
|\stackrel{1,0}{\Psi}{}_{\Phi}|^2 \stackrel{1,0}{\Psi}{}_{\Phi}
=0,
\end{eqnarray*}
$$
\stackrel{1,0}{\Psi}{}_{\Phi}|_{l=0}= \int_{-\infty}^{\infty} d\s
f(x_1)\exp(i\int_0^\s d\chi \l(x_1,t_1,\ve)),
$$
where the $\xi$ is defined from
$$
{{dx_1}\over{d\xi}}=\pt_{t_2}\Phi,\quad
{{dt_1}\over{d\xi}}=\pt_{x_2}\Phi.
$$
The coefficients $\stackrel{n,k}{\Psi}{}_{\pm\Phi}$ are determined
from Cauchy problems for linearized Schrodinger equation
(\ref{lSh}). The coefficients $\stackrel{n,k}{\Psi}_{\chi}$,
$\chi\in K'_{n,k}$ are determined from algebraic equations
(\ref{algebraic}). The set $K'_{n,k} = K_{n,k} \backslash \{\pm
\Phi \}$.
\end{theorema}
Theorem  1 follows from theorems 2,3 and 4.

\subsection{Structure of the second external asymptotics}
\par
Let us construct the formal asymptotic solution from theorem
\ref{theoremAboutSecondAsymptotics}. Substitute (\ref{exaf}) in
original equation and collect the terms of the same order with
respect to $\ve$. It yields $N+1$ equations and residual of the order
$\ve^{N+1}$. After collecting the terms with the same phase functions
we obtain the recurrent system of equation for coefficients of
(\ref{exaf}).
\par
Let us consider equations under $\exp(i\Phi/\ve^2)$. The terms of the
order $\ve^1$ give us the equation (\ref{eikonal2}) for the phase
function of eigen oscillations. The initial data is determined by
matching condition and represented by value of driven phase $S$ on
the resonance curve $l=0$
$$
\Phi|_{l = 0} = S|_{l=0},\quad
\pt_{t_2}\Phi|_{l=0}=\pt_{t_2}S|_{l=0}.
$$
The terms of the order $\ve^2$
$$
2i\left(\pt_{t_2}\Phi\pt_{t_1}\stackrel{1,0}{\Psi}{}_{\Phi} -
\pt_{x_2}\Phi\pt_{x_1}\stackrel{1,0}{\Psi}{}_{\Phi}\right) =0
$$
give us the homogeneous transport equation
\bb
\pt_{t_2}\Phi\pt_{t_1}\stackrel{1,0}{\Psi}{}_{\Phi} -
\pt_{x_2}\Phi\pt_{x_1}\stackrel{1,0}{\Psi}{}_{\Phi} =0.
\label{perenos}
\ee
This equation allows us to determine the dependence of the
leading-order term  on characteristic variable $\zeta$. Equation
(\ref{perenos}) along the characteristics
\bb
{{d x_1}\over{d \zeta}} = - \pt_{x_2}\Phi, \quad  {{d t_1}\over{d
\zeta}} =  \pt_{t_2}\Phi
 \label{charactereq}
\ee
can be written in the form of ordinary differential equation
\bb
{{d \stackrel{1,0}{\Psi}{}_{\Phi}}\over{d \zeta}}=0.
\label{eqInFastVariablesForPsi1}
\ee
It yields $\stackrel{1,0}{\Psi}{}_{\Phi}$ depends on $\xi$, where
the $\xi$ is defined by
$$
{{dx_1}\over{d\xi}}=\pt_{t_2}\Phi,\quad
{{dt_1}\over{d\xi}}=\pt_{x_2}\Phi.
$$
\par
The terms of the order $\ve^3$ which oscillates as
$\exp(i\Phi/\ve^2)$ are
\begin{eqnarray*}
2i\left(\pt_{t_2}\Phi\pt_{t_1}\stackrel{2,0}{\Psi}{}_{\Phi} -
\pt_{x_2}\Phi\pt_{x_1}\stackrel{2,0}{\Psi}{}_{\Phi}\right) + \nonumber\\
2i\pt_{t_2}\Phi\pt_{t_2}\stackrel{1,0}{\Psi}{}_{\Phi} +
[(\pt_{t_1}\xi)^2-(\pt_{x_1}\xi)^2]\pt^2_{\xi\xi}\stackrel{1,0}{\Psi}{}_{\Phi}
+ \\
 i[\pt_{t_2}^2\Phi - \pt_{x_2}^2\Phi]\stackrel{1,0}{\Psi}{}_{\Phi} +
\gamma |\stackrel{1,0}{\Psi}{}_{\Phi}|^2
\stackrel{1,0}{\Psi}{}_{\Phi} =0.
\end{eqnarray*}
It is convenient to write this equation  in the form of ordinary
differential equation  in terms of characteristic variables
\begin{eqnarray}
{{d \stackrel{2,0}{\Psi}{}_{\Phi}}\over{d \zeta}} =
-2i\pt_{t_2}\Phi\pt_{t_2}\stackrel{1,0}{\Psi}{}_{\Phi} -
[(\pt_{t_1}\xi)^2-(\pt_{x_1}\xi)^2]\pt^2_{\xi\xi}\stackrel{1,0}{\Psi}{}_{\Phi}
\nonumber\\
-  i[\pt_{t_2}^2\Phi -
\pt_{x_2}^2\Phi]\stackrel{1,0}{\Psi}{}_{\Phi} - \gamma
|\stackrel{1,0}{\Psi}^{\pm}|^2 \stackrel{1,0}{\Psi}{}_{\Phi}.
\label{laste}
\end{eqnarray}
Equation (\ref{eqInFastVariablesForPsi1}) shows  that the right hand
side of equation (\ref{laste}) does not depend on $\zeta$. To avoid
secularities in the asymptotics we  demand  the right hand side of
equation is equal to zero. It allows to determine the dependence of
the leading-order term  on slow variable $t_2$
\begin{eqnarray}
2i\pt_{t_2}\Phi\pt_{t_2}\stackrel{1,0}{\Psi}{}_{\Phi} +
[(\pt_{t_1}\xi)^2-(\pt_{x_1}\xi)^2]\pt^2_{\xi\xi}\stackrel{1,0}{\Psi}{}_{\Phi}
+  \nonumber\\
+i[\pt_{t_2}^2\Phi -
\pt_{x_2}^2\Phi]\stackrel{1,0}{\Psi}{}_{\Phi} + \gamma
|\stackrel{1,0}{\Psi}{}_{\Phi}|^2 \stackrel{1,0}{\Psi}{}_{\Phi}
=0. \label{nls}
\end{eqnarray}
\par
The equations for  the higher-order  terms  are obtained by  the same
manner
$$
2i\left(\pt_{t_2}\Phi\pt_{t_1}\stackrel{n+1,k}{\Psi}{}_{\Phi} -
\pt_{x_2}\Phi\pt_{x_1}\stackrel{n+1,k}{\Psi}{}_{\Phi}\right)
=2i\pt_{t_2}\Phi\pt_{t_2}\stackrel{n,k}{\Psi}{}_{\Phi} -
\pt^2_{\xi\xi}\stackrel{n,k}{\Psi}{}_{\Phi} -
$$
$$
-i[\pt_{t_2}^2\Phi -
\pt_{x_2}^2\Phi]\stackrel{n,k}{\Psi}{}_{\Phi} +
\pt_{t_1}\xi\pt^2_{\xi t_2}\stackrel{n-1,k}{\Psi}{}_{\Phi}- \gamma
\sum_{k_1,k_2,l_1,l_2,m_1,m_2,\alpha,\beta,\delta}
\stackrel{k_1,k_2}{\Psi}_{\alpha} \stackrel{l_1,l_2}{\Psi}_{\beta}
 \stackrel{m_1,m_2}{\Psi}_{\delta},
$$
where $k_1+l_1+m_1=n+2,\ k_2+l_2+m_2=k,\ \alpha + \beta + \delta =
\Phi,\ \alpha\in K_{k_1,k_2},\ \beta\in K_{l_1,l_2},\ \delta \in
K_{m_1,m_2}.$
\par
To construct the uniform asymptotic expansion with respect to
$\zeta$ we obtain the linearized Schrodinger equation for
higher-order  term
\begin{eqnarray}
2i\pt_{t_2}\Phi\pt_{t_2}\stackrel{n,k}{\Psi}{}_{\Phi} +
\pt^2_{\xi\xi}\stackrel{n,k}{\Psi}{}_{\Phi} +i[\pt_{t_2}^2\Phi -
\pt_{x_2}^2\Phi]\stackrel{n,k}{\Psi}{}_{\Phi} =
\nonumber \\
- \pt_{t_1}\xi\pt^2_{\xi t_2}\stackrel{n-1,k}{\Psi}{}_{\Phi}
-\gamma \sum_{k_1,k_2,l_1,l_2,m_1,m_2,\alpha,\beta,\delta}
\stackrel{k_1,k_2}{\Psi}_{\alpha} \stackrel{l_1,l_2}{\Psi}_{\beta}
 \stackrel{m_1,m_2}{\Psi}_{\delta},\label{lSh}
\end{eqnarray}
where $k_1+l_1+m_1=n+2,\ k_2+l_2+m_2=k,\ \alpha + \beta + \delta =
\Phi,\ \alpha\in K_{k_1,k_2},\ \beta\in K_{l_1,l_2},\ \delta \in
K_{m_1,m_2}.$
\par
The amplitudes $\stackrel{n}{\Psi}_{\chi}$ as $\chi\not=\pm\Phi$ are
determined by algebraic equations
\bb
\left[-(\chi_{t_2})^2 + (\chi_{x_2})^2 + 1
\right]\stackrel{n,k}{\Psi}_{\chi} = \stackrel{n,k}F_{\chi},\quad
\chi\not= \pm \Phi \label{algebraic}.
\ee
Here the right hand side of the equation depends on previous
terms and their derivatives
$$
\stackrel{n,k}F_{\chi} =
-2i\chi_{t_2}\pt_{t_1}\stackrel{n-1,k}{\Psi}_{\chi} +
2i\chi_{x_2}\pt_{x_1}\stackrel{n-1,k}{\Psi}_{\chi} - 2i \chi_{t_2}
\pt_{t_2}\stackrel{n-2,k}{\Psi}_{\chi} - i\left[\chi_{t_2 t_2} -
\chi_{x_2 x_2} \right]\stackrel{n-2,k}{\Psi}_{\chi}-
$$
$$
 \pt^2_{t_1
t_2}\stackrel{n-3,k}{\Psi}_{\chi} - \pt^2_{t_2
t_2}\stackrel{n-4,k}{\Psi}_{\chi} -\gamma
\sum_{k_1,k_2,l_1,l_2,m_1,m_2,\alpha,\beta,\delta}
\stackrel{k_1,k_2}{\Psi}_{\alpha} \stackrel{l_1,l_2}{\Psi}_{\beta}
\stackrel{m_1,m_2}{\Psi}_{\delta},
$$
where $k_1+l_1+m_1=n-4,\ k_2+l_2+m_2=k,\ \alpha + \beta + \delta
= \chi,\ \alpha\in K_{k_1,k_2},\ \beta\in K_{l_1,l_2},\ \delta
\in K_{m_1,m_2}.$
\par
These equations are similar to equations for amplitudes from
pre-resonance section.
\par
The obtained result is formulated below
\begin{lemma}
\label{lemmaAboutEquationsForSecondExternalAsymptoticExpansion}
The coefficients of formal asymptotic solution (\ref{exaf})
satisfy recurrent system of equations (\ref{eikonal2}),
(\ref{nls}), (\ref{lSh}), (\ref{algebraic}).
\end{lemma}
\par
The right hand side of equation (\ref{lSh}) has a singularity as
$l \to 0$. The singularity appears  due to
$\stackrel{n,k}{\Psi}_{\chi}$ as $\chi\not=\pm\Phi$. The analysis
of the right hand side of the equation allows us to calculate the
order of singularity as $l\to 0$. It is equal to $O(l^{-(n-1)})$.
Below we prove the solvability of equation (\ref{lSh}) with the
given asymptotics as $l\to 0$.
\begin{lemma}\label{lemmaAboutAsymptoticsForLS}
The asymptotics as $l \to 0$ of the solution of equation (\ref{lSh})
has the  form
\bb
\stackrel{n,k}{\Psi}{}_{\Phi}(x_1,t_1,t_2) =
\sum_{j=-(n-2)}^{1}\sum_{m=0}^{j-1}
\stackrel{n,k}{\Psi}{}_{\Phi}^{j,m}(x_1,t_1)\ l^j(\ln l)^m + O(1),
\quad l\to 0. \label{asymptotics_for ls}
\ee
\end{lemma}
{\bf Proof.} Determine the order of the singularity of the right
hand side of the equation as $l \to 0$. First consider  equation
(\ref{lSh}) for $n=3, k=0$. The solution of this equation gives us
the coefficient $\stackrel{3,0}{\Psi}{}_{\Phi}$. The nonlinearity
contains the term  $|\stackrel{2,0}{\Psi}_{S}|^2
\stackrel{1,0}{\Psi}_{\Phi}$. The function
$\stackrel{2,0}{\Psi}_{S}$ has the singularity of the order
$l^{-1}$ as $l \to 0$. It determines the order of singularity for
right hand side $l^{-2}$. We construct the asymptotics of
$\stackrel{3,0}{\Psi}{}_{\chi}$ in the form
\bb
\stackrel{3,0}{\Psi}{}_{\Phi} =
\stackrel{3,0}{\Psi}{}_{\Phi}^{-1,0} l^{-1} +
\stackrel{3,0}{\Psi}{}_{\Phi}^{0,1} \ln(l) +
\stackrel{3,0}{\Psi}{}_{\Phi}^{1,1} l \ln(l) + \widehat{
\stackrel{3,0}{\Psi}}{}_{\Phi}, \label{asym_n3}
\ee
Substitute (\ref{asym_n3}) in equation for $n=3$. It leads to
recurrent system of equations for coefficients
$\stackrel{3,0}{\Psi}{}_{\Phi}^{(j,k)}$
$$
-2i \pt_{t_2}\Phi\pt_{t_2}l\stackrel{3,0}{\Psi}{}_{\Phi}^{(-1,0)}
= - \stackrel{1,0}{\Psi}_{\Phi} |\stackrel{2,0}{\Psi}{}_{S}|^2
l^2,
$$
$$
2i\pt_{t_2}\Phi\pt_{t_2}l\stackrel{3,0}{\Psi}{}_{\Phi}^{(0,1)}=
L[\stackrel{3,0}{\Psi}{}_{\Phi}^{(-1,0)}],
$$
$$
2i\pt_{t_2}\Phi\pt_{t_2}l\stackrel{3,0}{\Psi}{}_{\Phi}^{(1,1)}=
L[\stackrel{3,0}{\Psi}{}_{\Phi}^{(0,1)}].
$$
Here we denote the linear operator by
$$
L[\Psi]=2i\pt_{t_2}\Phi\pt_{t_2}\Psi+\pt_\xi^2\Psi+i[\pt_{t_2}^2\Phi-\pt_{x_2}^2\Phi]\Psi+
\gamma\big(2|\stackrel{1,0}{\Psi}_\Phi|^2\Psi+
(\stackrel{1,0}{\Psi}_\Phi)^2\Psi^*\big).
$$
\par
The regular part $\widehat{\stackrel{3,0}{\Psi}{}_{\Phi}}$ of the
asymptotics satisfies the nonhomogeneous linear Schrodinger
equation. The right hand side of the equation is smooth
$$
L[\widehat{
\stackrel{3,0}{\Psi}}{}_{\Phi}]=-l\ln|l|L[\stackrel{3,0}{\Psi}{}^{(1,1)}_\Phi]-
2i\pt_{t_2}\Phi\pt_{t_2}l \stackrel{3,0}{\Psi}{}^{(1,1)}_\Phi.
$$
The initial condition for the regular part of the asymptotics is
determined below by matching with the internal asymptotic expansion.
\par
The structure of the terms $\stackrel{n,k}{\Psi}_{\pm\Phi}$ for
$n>3$ has a similar form. The right hand side of equation
(\ref{lSh}) depends on junior terms. These singularities can be
eliminate
$$
\stackrel{n,k}
F_{\Phi}=\sum_{j=0}^{-(n-2)}\sum_{m=0}^{-j+1}l^j\ln^m|l|
\stackrel{n,k}{f}{}^{(j,m)}_{\Phi}+\widehat{
\stackrel{n,k}{F}}_{\Phi}.
$$
The coefficients $\stackrel{n,k}{f}{}^{(j,m)}_{\Phi}$  do not contain
singularities as $l\to 0$. These coefficients are easy calculated.
\par
The direct substitution of (\ref{asymptotics_for ls}) in  equation
and collecting the terms with the same order of $l$ complete the
proof of lemma \ref{lemmaAboutAsymptoticsForLS}.

\subsection{The domain of validity of the second external asymptotics
and matching procedure}
\par
The domain of validity of the second external asymptotics is
determined by
$$
{\ve\stackrel{n+1}{V}\over\stackrel{n}{V}}\ll1.
$$
Formulas (\ref{exaf}) and  (\ref{asymptotics_for ls}) give the
condition
$$
l\ll\ve.\label{validityOfSecondExtAs}
$$
\par
The domain $|l|\ll1$ of validity of the internal asymptotics and
domain of validity of the second external asymptotics are
intersected. This fact allows to complete  the construction of the
second external asymptotics by matching method \cite{Il'in}.  The
structure of singular parts of the internal asymptotics as  $\l\to
+\infty$ and external asymptotics as $l \to 0$ are equivalent. The
coefficients are coincided due to our constructions. The matching of
regular parts of these asymptotics takes place due to
$$
\stackrel{n,0}{\Psi}_{\Phi}|_{l=0}=\stackrel{n}{W}{}^{(0,0)}(\xi).
$$
The function  $\stackrel{n}{W}{}^{(0,0)}(\xi)$ is determined in lemma
\ref{lemmaAboutAsymptoticsAsLambdaToInfonity}.
\par
In particular, the initial condition for the leading-order term has
a form
$$
\stackrel{1,0}{\Psi}{}_{\Phi}|_{l=0}= \int_{-\infty}^{\infty} d\s
f(x_1)\exp(i\int_0^\s d\chi \l(x_1,t_1,\ve)).
$$
The soliton theory for nonlinear Schrodinger equation leads us to
the fact that the function $\stackrel{1,0}{\Psi}{}_{\Phi}$ contains
the solitary waves when $f(x_1)$ is sufficiently large.
\par
Theorem  \ref{theoremAboutSecondAsymptotics} is proved.
\par
{\bf Acknowledgments.} We are grateful to  I.V. Barashenkov, L.A.
Kalyakin and B.I.Suleimanov for helpful comments and for help in
improving of the mathematical presentation the results.

\end{document}